\def\({\left(}
\def\[{\left[}
\def\l\{{\left\{}
\def\){\right)}
\def\]{\right]}
\def\r\}{\right\}}
\def\what{\widehat}
\def\raw{\rightarrow}
\def\bA{\bar A}
\def\am{\bar A_{\rm max}}
\def\cf{{\cal F}}

\def\etal{{\sl et al.} }
\def\ratio#1#2{{{#1}\over{#2}}}

\def\DXDYCZ#1#2#3{\left({\partial#1\over\partial#2}\right)_{#3}}
\def\lapp{\mathbin{\raise2pt \hbox{$<$} \hskip-9pt \lower4pt
\hbox{$\sim$}}}
\def\gapp{\mathbin{\raise2pt \hbox{$>$} \hskip-9pt \lower4pt
\hbox{$\sim$}}}
\documentclass{aa}
\usepackage{graphicx}
\begin{document}

   \thesaurus{02.01.1; 02.08.1; 02.09.1; 02.19.1; 11.10.1}
   \title{Diffusive shock 
acceleration in extragalactic jets}

   \author{ M. Micono\inst{1}
   \and N. Zurlo\inst{2} \and S. Massaglia\inst{1} \and
    A. Ferrari\inst{1} \and D.B. Melrose\inst{3}
          }

\institute{Dipartimento di Fisica Generale dell'Universit\`a,
Via Pietro Giuria 1, I-10125 Torino, Italy \and
Dipartimento di Ingegneria Aeronautica e Spaziale, Politecnico di Torino,
Corso Duca degli Abruzzi 24, I-10129 Torino, Italy \and
Research Centre for Theoretical Astrophysics,
School of Physics, University of Sydney, NSW 2006, Australia}

   \offprints{M. Micono}

   \date{Received; accepted }

\maketitle

\begin{abstract}
We calculate the temporal evolution of distributions
of relativistic electrons subject to synchrotron and adiabatic
processes and Fermi-like acceleration in shocks. The shocks result
from Kelvin-Helmholtz instabilities in the jet. Shock formation
and particle acceleration are treat\-ed in a self-consistent way
by means of a numerical hydrocode.
We show that in our model the number of relativistic particles is
conserved during the evolution, with no need of further injections
of supra-thermal particles
after the initial one. From our calculations, we
derive predictions for values and trends
of quantities like the spectral index and the cutoff frequency
that can be compared with observations.

   \end{abstract}
\section{Introduction}
Observations of non-thermal astrophysical sources upon
the utmost diverse spatial scales and luminosity ranges, from Supernova
remnants to Active Galactic Nuclei, and energy bands,
from radio frequencies up to $\gamma-$rays,
require electron acceleration
up to ultra-relativistic energies when interpreted via synchrotron or
self-Compton emission. Relativistic particles can also be directly
detected as in the case of the streams of particles from solar
flares or the cosmic radiation.

Multi-wavelength studies of extragalactic jets, per\-for\-med by
combining observations from radio arrays, HST in the optical band
and X-ray satellites, showed
that many extragalactic jets emit non-thermal radiation
extending from radio to X-rays,
and that spectral and morphological features remain,
in many instances, nearly constant
along the jet from radio up to optical frequencies. This implies a
constancy of the relativistic electron distribution function over at
least five decades in energy, that needs to be interpreted (see
discussions in Meisenheimer, R\"oser \& Schl\"otelburg 1996 and
Meisenheimer, Neumann \& R\"oser 1996 for M87).

Many physical mechanisms of particle acceleration 
have been studied, and Fermi-like processes occurring at MHD shocks
have been particularly visited since they  automatically lead
to power-law particle distributions, as dictated by observations
(see, e.g.,
Bell 1978, Drury 1983, Blandford \& Eichler
1987, Achterberg 1990, Kirk 1994).
Other mechanisms of particle acceleration include
 the diffusive particle acceleration at a tangential
discontinuity in a velocity field involving relativistic velocities
(Ostrowski, 1990, 1998). Continuous reacceleration of particles
could take place by conversion of magnetic energy into particle
energy via magnetic reconnection, as proposed by
Lesch \& Birk (1998); mechanisms of particle acceleration by
turbulent plasma waves (Li et al. 1997) or by intense long-wavelength
electromagnetic fields (Bisnovatyi-Kogan \& Lo\-ve\-la\-ce 1995) have also been
considered. Each of these mechanisms leads to different values for the
maximum energy attained by the accelerated particles, and different
predictions on the observed radiation spectrum.

A possible clue to build a ``universal spectrum"
is the  diffusive shock acceleration (DSA) by
multiple sh\-ocks, as discussed in Melrose \& Pope (1993) and Ferrari \&
Melrose (1997), under very general conditions.

In this paper we concentrate on the diffusive particle acceleration
at shocks, and we
treat, in a consistent way by means of a
numerical hydrocode, the jet instabilities that yield shock formation,
the particle acceleration in these shocks, and the temporal evolution of
the distribution function, subject to adiabatic effects and
synchrotron losses. 
In our model the magnetic field and the
relativistic particles are advected passively by the thermal fluid,
as in previous studies by Matthews and Scheuer (1990) who treated 
adiabatic expansion and synchrotron losses without including 
particle acceleration, and by Massaglia et al. (1996) who modeled shock
acceleration as an adiabatic compression of fixed strength. In our study
the equation for the evolution of the particle distribution function is 
solved in a self-consistent way, and particle acceleration 
at shocks is treated in the `test-particle'
approximation, i.e. neglecting the effects of energy subtraction from the 
shocks.
Moreover we concentrate on the 
stability and radiative properties of the jet beam, rather
than studying the propagation of the jet's head as it is done in the
quoted papers. Our approach is thus also different from that of Jones et al.
(1999) who solve a simplified version of the electron transport equation 
to account for shock diffusive acceleration and synchrotron aging in a 
time-dependent simulation of a radio lobe.
 
In studying the instability evolution, we adopt
the so-called `temporal' approach (see Bodo et al. 1994), in which one
can follow the system evolution for longer times, as opposed to the
`spatial analysis' (see Hardee \& Norman 1988a,b and Stone, Xu \&
Hardee 1997),  in which one is limited by  the transit time through
the grid. While  the spatial approach allows for more direct
comparison with observations, the temporal analysis is preferred
when studying the physics of the instability evolution over long time
scales, with good spatial resolution and without exceedingly large
computational domains.

The plan of the paper is the
following: in the next section (Section 2), we  describe the
physical model for the jet, the main
assumptions and  the integration method; in Section 3 we deal
with the treatment of the evolution
equation for the relativistic electrons distribution
and of the shock acceleration;
 the simulations results are discussed in Sections 4  and 5
where a comparison between our model and observational data
of extragalactic jets is presented, 
and  the conclusions are given in Section 6.

\section{The jet physical model}
We consider a non-relativistic, fluid jet that propagates in a uniform
medium and is in pressure equilibrium with its environment. This
environment is permeated by a {\it  passive} magnetic field, that is
advected by the fluid and has no  effect in the  momentum conservation
equation ($\beta_{\rm plasma} \rightarrow \infty$).
Under these conditions, the relevant equations are the hydrodynamic
equations of mass, momentum and energy conservation:
\begin{equation}
\frac{\partial \rho}{\partial t} + \nabla \cdot (\rho
\vec{v}) = 0 
\end{equation}
\begin{equation}
\frac{\partial \vec{v}} {\partial t} + (\vec{v} \cdot
\nabla)\vec{v} = -\nabla p / \rho 
\end{equation}
\begin{equation}
\frac{\partial p} {\partial t} + (\vec{v} \cdot \nabla)p -
\Gamma \frac{p} {\rho} \left[
\frac{\partial \rho} {\partial t} + (\vec{v} \cdot \nabla) \rho
\right] =0 \enspace, 
\end{equation}
where, as usual, $p$ represents the thermal pressure, $\rho$
the density, ${\bf v}$ the fluid velocity,
and $\Gamma$ stands for the ratio of specific heats.

We then restrict our analysis to an
infinite jet in cylindrical geometry (in the coordinates $r,z$);
consistently with our assumptions, the equation system (1),(2),(3) can be
complemented by the equations for the passive magnetic field,
in the form:
\begin{equation}
\frac{\partial}  {\partial t}  (r A_{\phi}(r,z))  + \vec{v}
\cdot \nabla (r A_{\phi}(r,z) )  = 0        
\label{bfield1}    
\end{equation}
\begin{equation}
\frac{\partial} {\partial t} \left( B_{\phi}(r,z) \over \rho r \right) +
\vec{v}  \cdot \nabla
 \left( B_{\phi}(r,z) \over \rho r \right)
 = 0
\,,
\label{bfield2}
\end{equation}
where $A_{\phi}(r,z)$ is the only component of the vector potential
for $B_z$, $B_r$
($r A_{\phi}(r,z)$ is usually called `stream function'), as appropriate
for the chosen geometry, and $B_{\phi}(r,z)$ is the toroidal field.
One can notice that Eqs. (4,5) have in common the
standard form of a `tracer' equation:
\begin{equation}
\frac{\partial {\cal T}} {\partial t} + \vec{v}  \cdot \nabla
  {\cal T}  = 0 \enspace, \label{tracer}
\end{equation}
where ${\cal T} \equiv r A_{\phi}$ in Eq.~(\ref{bfield1})
and ${\cal T} \equiv
B_{\phi} / \rho r$ in Eq.~(\ref{bfield2}).
\subsection{Initial configuration and boundary conditions}
As mentioned before, we consider an axially symmetric, cylindrical jet.
The flow velocity is initially uniform
along the $z$ direction ($V_z$) and the jet is in pressure equilibrium
with the ambient medium. The initial velocity and density profiles in the $r$
coordinate, and the perturbation to the
transverse velocity $V_r(r,z)$ are the same as in Rossi et al. (1997).

In the calculations we measure lengths in units of the jet initial
radius $a$, time in units of the radius sound crossing time
$t_{\rm c} \equiv a/c_{\rm si}$ ($c_{\rm si}$ is the isothermal
sound speed internal to the jet), and the
magnetic field in units of the initial value $B_0$.

The initial configuration for the magnetic field is assumed
poloidal (longitudinal) plus azimuthal:
\begin{equation}
B_z=1 \,,\; B_r=0 \,,\; B_{\phi} = \frac{2 r}{\cosh(r^m)} \,,
\end{equation}
with $m=8$.

The system of equations (1),(2),(3) is solved numerically by means of a
PPM (Piecewise Parabolic Method) hydrocode (Colella \& Woodward 1984)
over an integration domain of $512 \times 256$ grid points,
with the jet radius spanning over 60 grid points, for a
total domain of $0 \leq z \leq 10 \pi$, $0 \leq r \leq 20$. The axis
of the jet  is coincident with the bottom boundary  of the domain
($r=0$), where symmetric (for $p$, $\rho$, $V_z$ and
$B_{\phi} / (\rho r)$) or  antisymmetric
(for $V_r$ and $r A_{\phi}$) boundary conditions are given.  At the $z=0$
and $z=20$ boundaries, consistently with the assumption
of a infinite jet, we have set periodic conditions,
and at the upper
boundary ($r=R$) we have chosen a free outflow condition, by imposing
for each variable null gradient ($d/dr=0$). The grid in the $r$
direction is non-uniform, but expands with $r$ to avoid the effects
of partial reflections at the top boundary (see Bodo et al. 1994
and Rossi et al. 1997).
\section{Evolution of relativistic electrons}

We study the evolution of a distribution of relativistic electrons 
as it is passively
advected by the fluid. The relativistic electrons are assumed to be
injected, at the initial time $t=0$,  with a 
power-law distribution
$N(E,0)= N_0 E^{-\gamma}$, defined on an  initial 
energy interval $10^{-4} < E/E_0 < 10^2$,
where $E_0$ will be defined below.
As the evolution proceeds, the electrons energy can vary on a  
wider interval $10^{-6} < E/E_0 < 10^4$.
A would-be representative sample of $n=10$  {\it parcels} of this
distribution are then selected and followed, as {\it Lagrangian test
particles},
as they travel with the fluid, undergoing adiabatic
expansion/compression effects, synchrotron losses and shock
acceleration.

\subsection{Energy evolution equation}
The temporal evolution equation for a  distribution function $N(E,t)$
for the number of relativistic electrons present in a parcel, 
subject to adiabatic
effects and synchrotron losses, can by written as (Kardashev 1962):
\begin{equation}
\frac{\partial N}{\partial t} = \frac{\partial}{\partial E}
\left[ \left( \alpha (t) E + \beta (t) E^2 \right) N \right] \enspace,
\label{kard}
\end{equation}
where
\begin{equation}
\alpha (t) = \frac{1}{3} \ \nabla \cdot \vec{v} \,,
\end{equation}
takes into account adiabatic effects, and
\begin{equation}
\beta (t) = b B^2 \,,
\end{equation}
accounts for synchrotron losses. Electron energy is expressed in units  of
the electron energy lost by synchrotron emission in a time unit in
the initial magnetic field,  $E_0=1/(b B_0^2 t_{\rm c})$.

Eq.~(\ref{kard}) is solved numerically for the selected test
par\-cels, away from shocks    
and at every time step, using a Lax-Wendroff scheme,
and the coupling between Eq.~(\ref{kard}) and the
hydrodynamic equations
(1) is given by the  expansion/compression term $\nabla \cdot \vec{v}$,
while the coefficient for synchrotron emission comes from the equations
for the magnetic field evolution (4) and (5).
The solution of the energy evolution equation is connected to the
hydrodynamic evolution through a time-splitting technique, with an
accurate control on the two Courant times.

We note that Kardashev (1962) gives an analytical solution for
Eq. (\ref{kard}) under the hypothesis that the initial 
distribution function is a power-law.
As time elapses, synchrotron losses and adiabatic effects modify the
shape of the distribution function in the high and low energy 
ranges of the spectrum respectively, and therefore the distribution
function evolves in time departing from a power-law.
In absence of shocks, one could adopt the Kardashev solution at any
time, 
but shocks introduce discontinuities in the evolution;
when the parcel has crossed a shock wave, the new initial
condition for the integration of Eq. (\ref {kard}) is the 
post-shock distribution function, which is not a power-law any longer 
and thus does not fulfill the Kardashev (1962) prescriptions.

\subsection{Diffusive shock acceleration}
When the selected test particle, representing a parcel of
the distribution, enters a shock, Fermi acceleration takes place.
Under the assumptions that the acceleration time scale $t_{\rm
acc} \ (\sim \kappa / v^2,$ with $\kappa$ the spatial diffusion
coefficient) is much smaller than
both the synchrotron time $t_{\rm sync}$, and the dynamical time
for shock evolution ($\sim t_{\rm c}$), we can apply the stationary
diffusive shock acceleration (DSA) model 
(see for example Drury 1983 for a review).
These conditions are fully verified in the energy interval over which
our electron distribution is allowed to vary, as it was defined 
above.

According to this model, the downstream number
of particles distribution $N_+(E,t)$ is
related to the upstream distribution $N_-(E,t)$ by:
\begin{equation}
N_+(E,t) = \frac{q}{r} \ E^{-q+2} \ \int_{E_{\rm min}}^E \
N_-(E^{\prime},t) \
E^{\prime \ q-3}
\ dE^{\prime} \enspace, \label{nplus}
\end{equation}
with
\begin{equation}
 q=\frac{3r}{r-1} \,,
\end{equation}
where $r=\rho_+ / \rho_-$ is the shock compression ratio,
 and $E_{\rm min}$ is the minimum electron energy in the 
pre-shock distribution function.
This equation is derived from the equation for the particle
density distribution function $f(E,t)$ (see for example Kirk 1994),

\begin{equation}
f_+(E,t) = {q} \ E^{-q+2} \ \int_{E_{\rm min}}^E \
f_-(E^{\prime},t) \
E^{\prime \ q-3}
\ dE^{\prime} \,,
\end{equation}

and the substitution $N(E,t)dE = f(E,t) V(t) dE$, 
where $V(t)$ is the volume occupied
by the particles, yields the compression ratio $r$ at
the denominator of Eq. (11).

We note that the post-shock distribution function is not a power-law
any longer.

White (1985), Achterberg (1990), Schneider (1993)
(see also the discussion in Melrose \& Pope 1993) found that
for an infinity of equally strong shocks
with $r=4$ and decompressions, the final distribution approaches
$N_+(E) \propto E^{-1}$.

\section{Results}

\begin{figure*}

{\includegraphics[width=\hsize,bb=10 150 700 740]{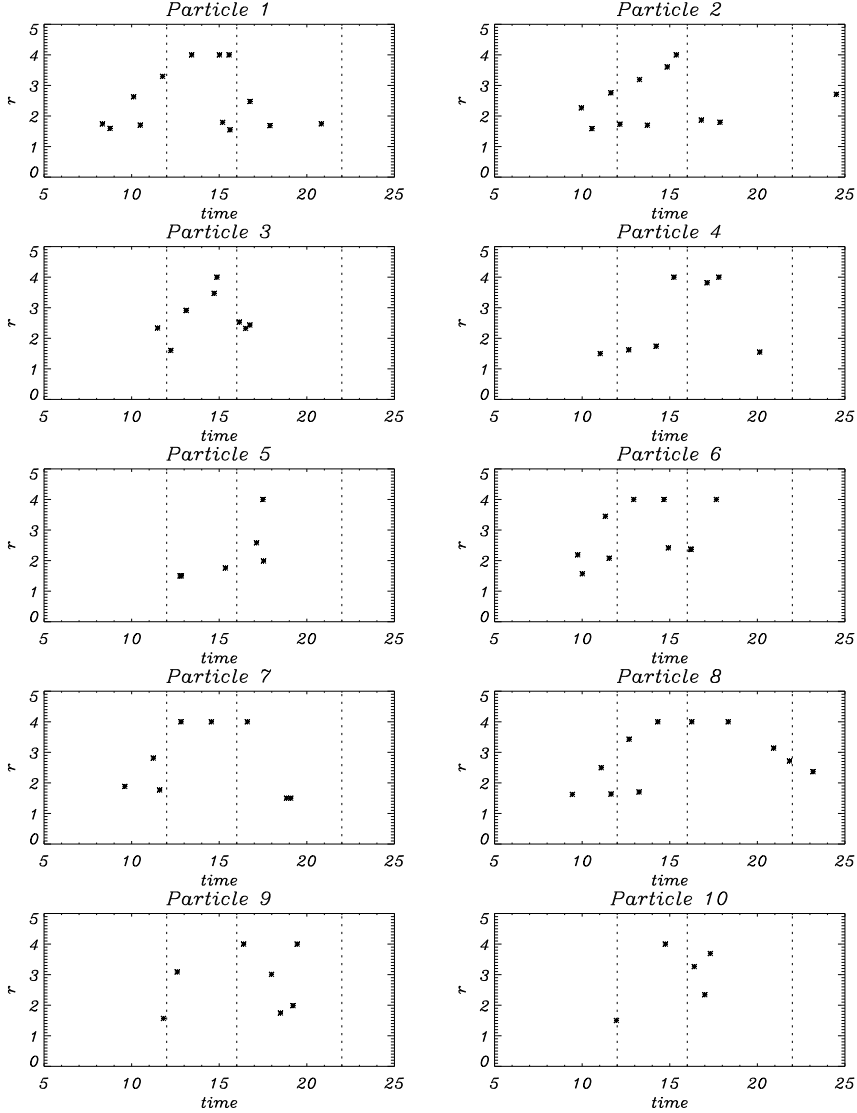}}

\caption{Compression ratio of each shock crossed by the Lagrangian 
particles as a function of time. The dashed lines are traced at the three
typical times $t=12$, $t=6$ and $t=22$ selected for the analysis and 
representative of different stages in the evolution. From this figure
the number and strength of the shocks can be inferred.}

\end{figure*}
             

%
We have carried out a simulation with a fluid jet defined by the
following parameters: Mach number
${\cal M} (\equiv v_z / c_{\rm si} \sqrt{\Gamma}) = 5$, density
ratio $\nu (\equiv \rho_{\infty} / \rho_{\rm jet}) =5$.
We studied the evolution of the distribution functions
associated with 10 Lagrangian test particles, whose initial positions
are given
in Table 1; the initial spectral index for the distribution functions
was $\gamma = 3.5$.  Consistently
with the assumed boundary conditions, particles
leaving the domain on the right boundary at $z=10\pi$ are re-injected at
$z=0$, and those leaving the domain at the upper boundary $r=20$ are lost.

\begin{table}[hbt]

\caption{Initial positions of the particles in the domain.}

\begin{center}
\begin{tabular}{|c|c|c|}  \hline

particle&$z$&$r$     \\
\hline
1      &12.21               &0.325  \\
2      &$\prime \prime$            &0.5  \\
3      &$\prime \prime$            &0.66  \\
4      &$\prime \prime$            &0.825  \\
5      &$\prime \prime$            &0.99  \\
6      &22.03        &0.24  \\
7      &$\prime \prime$            &0.41  \\
8      &$\prime \prime$            &0.575  \\
9      &$\prime \prime$            &0.74  \\
10      &$\prime \prime$            &0.91  \\
\hline
\end{tabular}
\end{center}
 \end{table}

To extract some physical quantities from our results  for a comparison
with the observational data, we needed to fix the scale quantities,
and these are summarized in Table 2.
They have been selected within the ranges
fixed by Birkinshaw (1991) for the
parameters characterizing radio jets.

\begin{table}[hbt]

\caption{Scaling parameters.}

\begin{center}
\begin{tabular}{|c|c|}  \hline

 & Simulation   \\
\hline
Magnetic field      & $B_0 = 3 \times 10^{-5}$ Gauss   \\
Jet radius          & a =  100 pc   \\
Sound velocity      & $c_{\rm s} = 10^8 {\rm \ cm \ s}^{-1}$  \\
Particles scale energy        & $E_0 = 0.21$ erg   \\
Particles Lorentz factor      & $\Gamma_0 = 2.55 \times 10^5 $   \\
\hline
\end{tabular} \end{center} \end{table}

\subsection{Instability evolution}

As described in detail in Bodo et al. (1994) the perturbed jet undergoes
three  stages of the instability:

\noindent
{\it Linear phase} (${\rm time} \le 7$): the
unstable modes excited by the perturbation grow in accord with
the linear theory. In the latter portion of this stage the growth of
the modes leads to the formation of internal shocks.

\noindent
{\it Acoustic phase} ($7 < {\rm time}\le 17$): the growth of  internal
shocks is accompanied by a global deformation of the jet;
energy and momentum are transferred from the jet to the external medium
through acoustic waves.

\noindent
{\it Mixing phase} (${\rm time} > 17$): as a consequence of the shock
evolution, mixing between jet and external material begins to occur.

\noindent
If the simulation is carried out further, a final
statistically quasi-stationary stage is attained.

\subsection{Particle spectrum evolution}

We now analyze the behavior of the Lagrangian particles and the
evolution of the distribution functions of the relativistic electrons
as these particles are advected through
the different stages of the instability evolution. We selected three
typical times,
$t=12$ in the middle of the acoustic phase, $t=16$ at the end, and $t=22$
when the mixing is already developed; the number of shocks crossed by
each particle at these given times, the average compression ratios
of these shocks and the power index of the distribution functions
are given in Table 3.
The details of the shock number and strength for each particle in each
phase can be inferred from Fig. 1.

\begin{table}[hbt]

\caption{Number of shocks crossed by a particle at a given time, and
exponent of the electrons distribution function}

\begin{center}
\begin{tabular}{|c|c|c|c|c|c|c|}  \hline

particle& \multicolumn{2}{|c|} {$t=12$} & \multicolumn{2}{|c|} {$t=16$} &
\multicolumn{2}{|c|}{$16<t<22$}     \\
\hline
 & n & $\gamma$ & n & $\gamma$ & n & $\gamma$ \\
\hline
1      & 5  &  2.3 & 10  &  1.6  & 13  & 3.6 \\
2      & 3  &  2.7 & 8   &  1.8  & 10  & 3.4 \\
3      & 1  &  3.2 & 6   &  1.8  & 9   & 2.1 \\
4      & 1  &  7.  & 4   &  2.   & 7   & 4.4 \\
5      & 0  &      & 3   &  5.   & 6   & 2.  \\
6      & 4  &  2.2 & 7   &  1.8  & 10  & 1.6 \\
7      & 3  &  2.6 & 5   &  1.9  & 8   & 2.4 \\
8      & 3  &  3.2 & 6   &  2.   & 11  & 1.6 \\
9      & 1  &  6.2 & 2   &  2.4  & 7   & 1.8 \\
10     & 1  &  7.  & 2   &  2.   & 5   & 1.9 \\
\hline
$r$    & \multicolumn{2}{|c|} {$1.5 - 2.5$} & \multicolumn{2}{|c|} {$ \sim
4$} & \multicolumn{2}{|c|} {$2.5 - 1.5$} \\
\hline

\end{tabular} \end{center} \end{table}

\begin{figure*}

{\includegraphics[width=\hsize,bb=10 10 700 740]{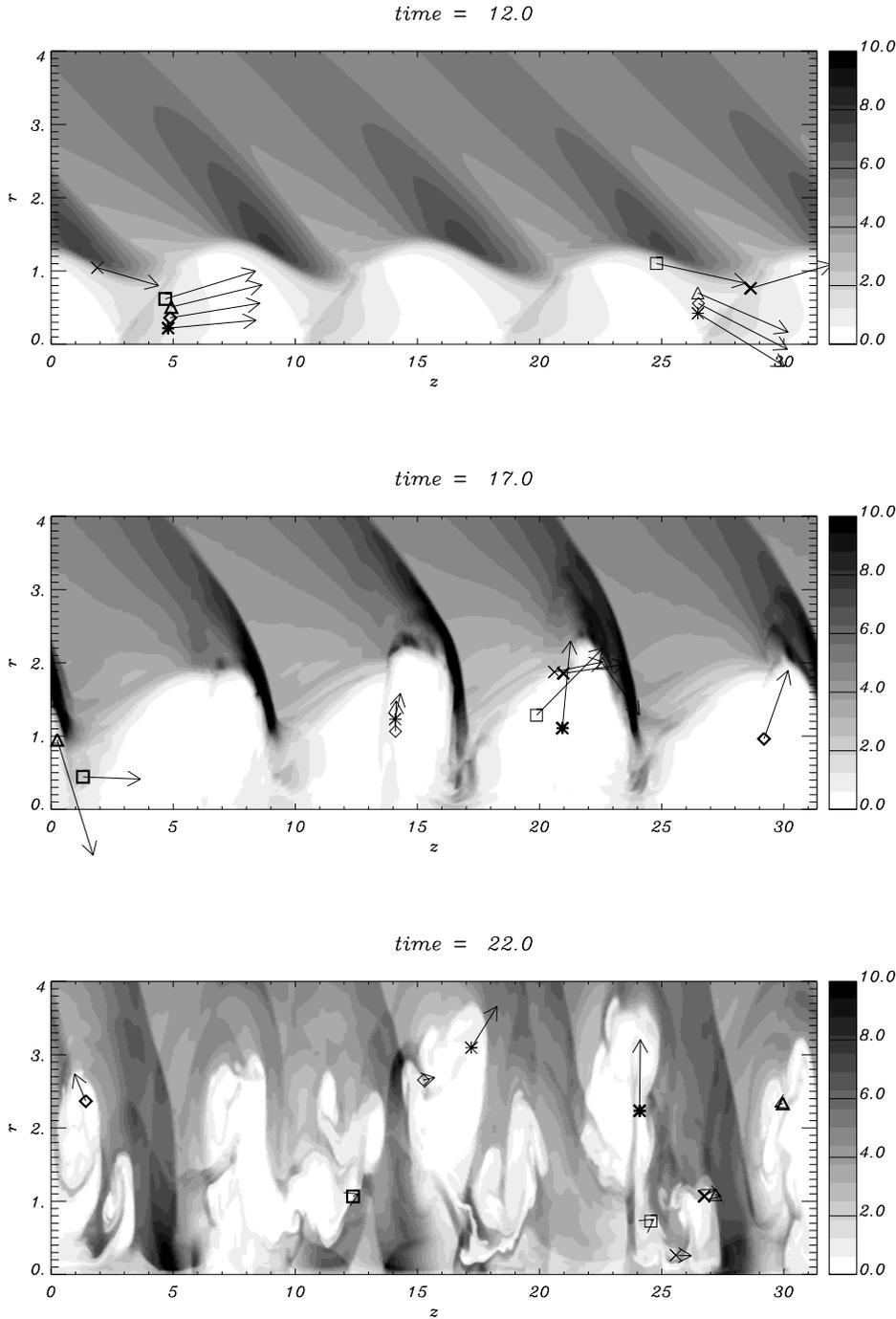}}

\caption{Grey-scale images of the jet density at three selected
times. The jet axis is located on the lower horizontal boundary
of the domain. The symbols superposed to the images show the positions
of the 10 Lagrangian particles, and the vectors represent their velocity.}

\end{figure*}

\begin{figure*}

{\includegraphics[width=\hsize, bb=10 500 700 740]{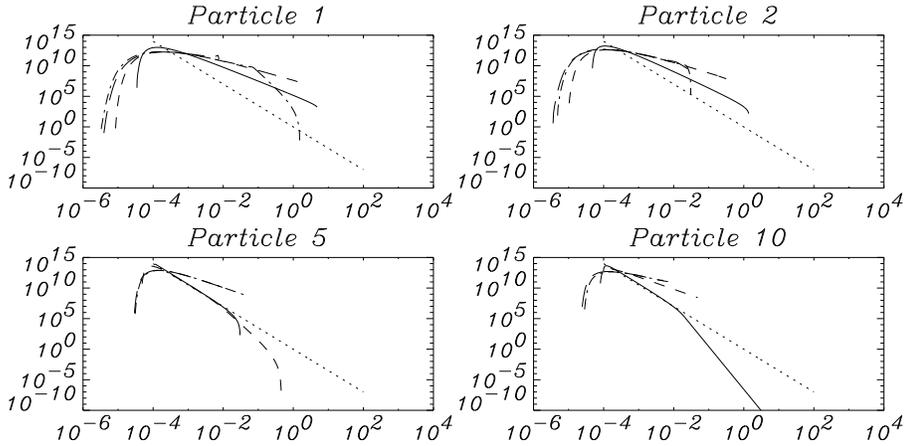}}

\caption{Distribution functions of the electrons associated to the
Lagrangian particles 1,2,5 and 10, at three selected times.
The dotted line represents the initial distribution function, the solid
line is for $t=12$, the dashed line for $t=16$ and the dash-dotted line
is for $t=22$.}

\end{figure*}

\begin{figure*}

{\includegraphics[width=\hsize,height=14.cm,bb=10 150 700 740]{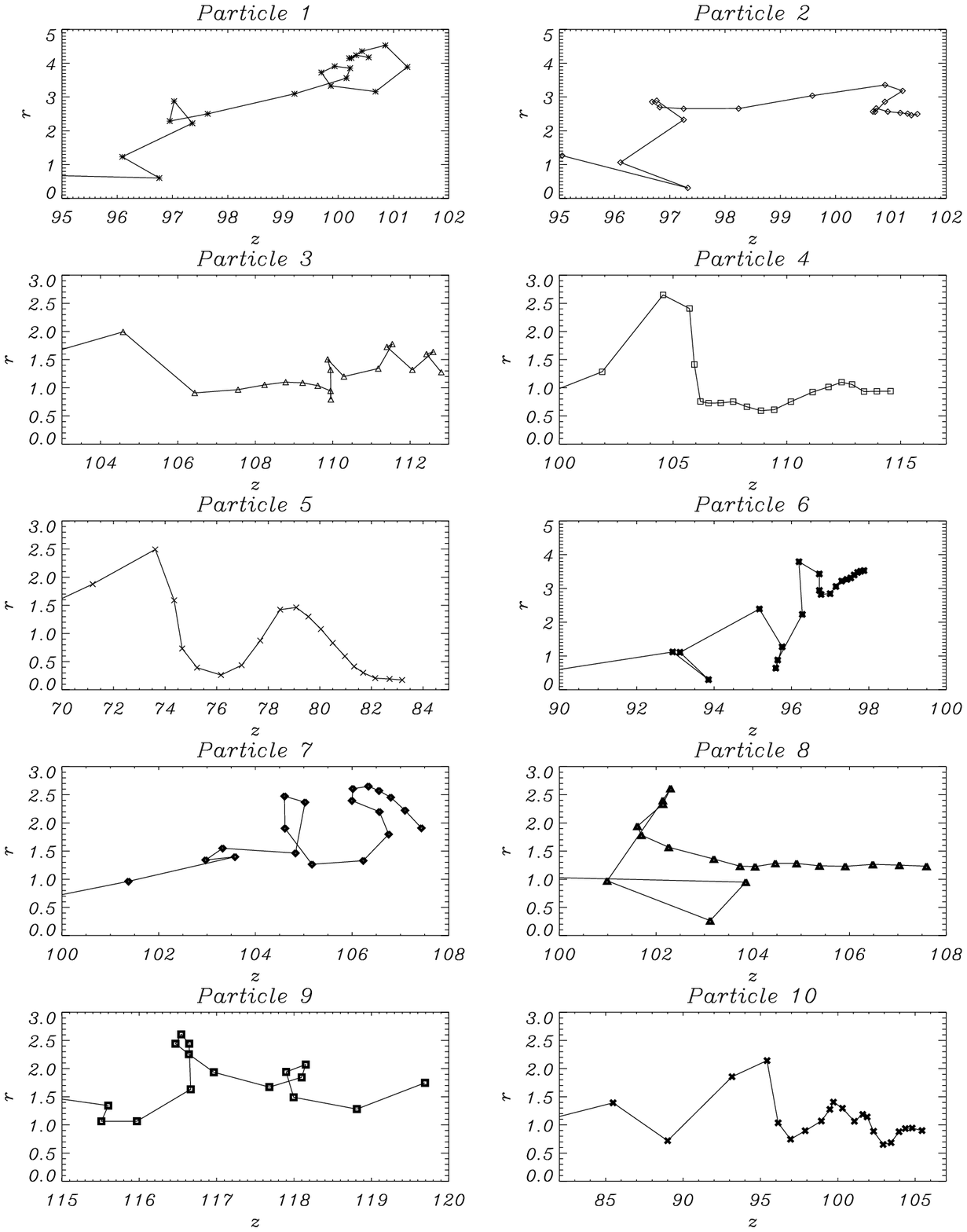}}

\caption{Trajectories of the ten particles plotted as a function
of distances from the injection point. Only the final sections of 
the trajectories are represented, since in previous stages the particles 
travel along a direction almost parallel to the jet axis, at a distance 
equal to the initial transversal position.}

\end{figure*}

In Fig. 2 we show three images of the jet density, in a logarithmic 
grey-scale, at these selected times, with
superimposed the positions
of the Lagrangian particles, identified by different symbols. The vectors
give the module and directions of the particles' velocities.
In Fig. 3 we plot the distribution functions associated to some of
the Lagrangian particles at the initial time and after the acceleration and loss
processes, at the same times selected above.

During the
linear phase the particles advance advected by the fluid and  the
relativistic electrons
lose energy th\-rou\-gh adiabatic expansion and synchrotron radiation.
The distribution functions steepen and the maximum energy diminishes,
as can be seen, for example, from the spectrum associated to
particle 5 at time
$t=12$ (continuous line in Fig. 3), when this particle has
not crossed any shock yet.

When shocks start to form, electrons are accelerated. The first
acceleration process occurs for the distribution function associated
to particle 1 at time $t=8.33$.
At time $t=12$
only the particles initially located at a distance less
than 0.7 jet radii from the jet axis have crossed a significant
($\sim 4$) number of shocks; at this stage shocks are still weak, with
compression ratios not exceeding $r
\sim 2.5$, and the electrons are only weakly accelerated; from Table 3
we see that the distribution function is a power law 
slightly flatter than the initial one. The particles located 
on the jet edges initially, instead, cross one or two shocks at most,
and their spectrum is still very steep ($\gamma \sim 6$) due to the
losses occurred in the linear phase of the evolution.

At time $t=16$ the jet is at the end of the acoustic phase;
the Lagrangian particles have crossed a high number of shocks
(on average $\sim 6$,
up to 10 for particle 1). Many of these shocks are very strong, with
compression ratios $r
\sim 4$, and the acceleration process results very efficient as can
be inferred from the values of the spectral index, which lays
at this stage in the interval $1.5 \lapp \gamma \lapp 2$.

There are three main reasons why particles whose initial position is
located near to the jet axis undergo a higher number of shock
accelerations: first, their longitudinal velocity is higher with respect
to the particles located on the edges of the jet;
second, their trajectory is less affected by motions in the radial
direction, since, due to the assumed geometry, 
the value of the radial velocity approaches zero 
near to the jet axis (see also Fig. 2);
third, the shocks that form due to the non linear growth of
the symmetrical body modes of the Kelvin-Helmholtz instability are
biconical
shocks whose vertex is located on the jet axis, and so the distance
between
shocks is smaller near the axis than near the edges.

After time $t=16$ the jet material starts to mix with the external medium,
shocks become less frequent and weaker (compression ratios
 $r \sim 2$),
until time $t=22$ when the mixing is well developed;
shocks are encountered less frequently by the particles,
whose velocity is diminished with
respect to the initial velocity, see Fig. 2c.
The spectral index of the electrons distribution functions
is different for the various particles: those which
cross a
small number of shocks acquire a steep spectral index,
comparable to the initial one (ex. particles 1,2,4,7) while
those which cross a conspicuous number of shocks
show a flatter spectrum ($\gamma \sim 2$; compare the values in the
last column of Table 3 and Fig. 1).

As the evolution progresses further, the mixing process
between the jet and the external medium dominates; the jet velocity
decreases, while momentum is transferred to the external medium, and
shocks lose strength until they disappear
(see Fig. 2c). The motion of the particles
is no longer ordered, they can be trapped in vortices and
cross regions of high magnetic field where electrons lose energy via
synchrotron radiation with no acceleration mechanism at work.
In Fig. 4 we show the trajectory of
the  Lagrangian particles as a
function of the distance from the point of injection; observe the
`loops' that appear in their paths, and how they
move far away from the jet axis, as the jet itself expands and
entrains external medium. The 'dots' on the paths in Fig. 4 are
snapshots of the particles' position at fixed time intervals
(corresponding to one time-scale): see how
their velocity diminishes as they move further downstream
along the jet.
As we extend our study further in time, the electrons spectrum steepen
and the spectrum moves towards lower and lower energies, approaching
the non-relativistic limit.

An important consequence of our model is the conservation of the total
number of relativistic electrons. We calculated
the integral of the electrons distribution functions and found that
it does not vary of more than the 2\% of the initial value during the
different stages of the evolution. This implies that no further
injection  of relativistic particles after the initial one
at the jet's source is needed; a finite, relatively small, number of shocks is
able to reaccelerate the electrons even if their maximum energy
is lowered and their distribution function is steepened by
synchrotron and expansion losses.

\section{On observable quantities}

\begin{figure*}

{\includegraphics[width=\hsize,height=14.cm,bb=10 150 700 740]{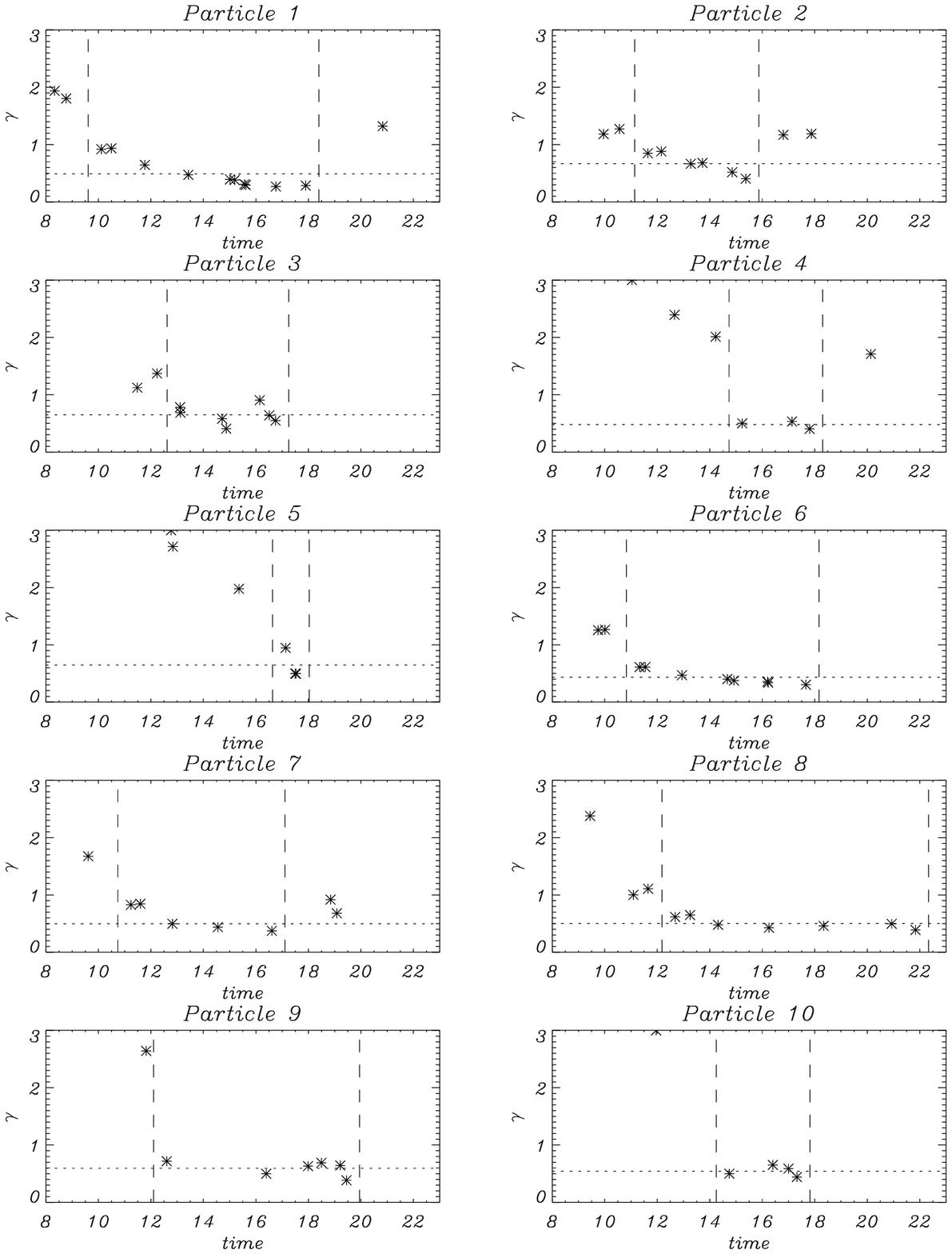}}

\caption{Post-shock radio (frequency interval: 
$10^9 \lapp \nu \lapp 6. \times 10^{11}$ Hz)
spectral index, at each shock crossed
by each Lagrangian particle.  The dotted lines correspond to the 
values obtained by averaging
over the shocks crossed in the time intervals defined by the 
dashed lines, and reported in the second column of Table 4.
 }
   \end{figure*}

From the results described in the previous section,
we can deduce some important values and trends
that can be directly compared to some observable quantities
characterizing extragalactic radio sources.
We compare the values obtained from our calculations to
those available for some well studied radio jets. It is important to 
notice that some of these sources have been recognized to be relativistic
(e.g. 3C 273, jet Lorentz factor $\Gamma_{\rm jet} \gapp 10$, Zensus et al. 1990) 
or weakly relativistic
(e.g. M 87, $\Gamma_{\rm jet} \lapp 2.5$, Reid et al. 1989, 
Biretta et al., 1995),
while our calculations are performed under the hypothesis that the 
bulk jet material is a non-relativistic thermal gas. We expect that
relativistic velocities introduce some important dynamical
 differences, regarding mainly the growth rates of 
the instability (Ferrari et al. 1978), and the properties of shock waves 
that form (Bicknell et al. 1996). Moreover, the standard diffusion
approximation that we adopted to study the particle acceleration
at shocks does not stand when shocks are relativistic: the anisotropy
of the particle distribution function must be taken into account and
the general solution of the transport equation must be constructed 
(Kirk \& Schneider 1987). In the relativistic case
the resulting spectrum of the accelerated particles has a strong dependence
on the shock compression ratio, and a small, though significant 
dependence on the form of the assumed diffusion coefficient 
(Heavens \& Drury 1988).
However if shocks that form are
oblique, as in the case of the Kelvin-Helmholtz induced biconical 
shocks, the {\it shock velocity} is reduced, with respect to the jet velocity,
by a geometrical factor; in this way the
{\it shock} Lorentz factor can be significantly smaller than the 
{\it jet} Lorentz factor; Heavens and Drury (1988) find that the 
non-relativistic treatment for particle acceleration at shocks
gives a reasonable estimate for the spectral index 
for shock Lorentz factors up to $\sim 1.15$: our results are
thus valid also for weakly relativistic jets, provided that
shocks in the flow are oblique, as it seems to be the 
case of knot A in M87 (Bicknell \& Begelman 1996).

Our results allow us to
calculate the spectral index of the radiation
emitted by the synchrotron mechanism, in different frequency intervals.
In Table 4 we show the values of
the radiation spectral index for the 10 particles
in those frequency ranges that are
relevant for observations, i.e. radio, radio-optical and optical-UV.
The values in Table 4 are average values of the spectral index
calculated by fitting the post-shock electrons spectra with a power law,
and averaging the values in a temporal interval corresponding
to the shock-dominated phase of the jet's evolution.

At radio frequencies,  we find that the spectral index varies 
in the range $0.43 \lapp \alpha_r \lapp 0.67$, consistently with the
values found in the majority of radio jets (see for example 
Bridle \& Perley 1984).

Only a few sources can be detected at optical frequencies, and thus the
data on radio to optical spectral index are available only for M87
($0.61 \le \alpha_{ro} \le 0.76$, Sparks et al.1996), 3C273
($\alpha_{ro} \sim 0.82$, Meisenheimer at al. 1996),
3C31 ($\alpha_{ro} \sim 0.8$, Fraix-Burnet et al. 1991a),
3C66B ($0.60 \le \alpha_{ro} \le 0.75$, Fraix-Burnet et al. 1991b),
and for other three cases (3C 147, 3C 239 and 3C 433) which show a 
radio to optical spectral index significantly higher.
From Table 4, the average spectral
 indexes associated to our particles in the radio to optical
frequency range ($10^9 \le \nu \le 7.7 \times 10^{14}$ Hz) are consistent
with the first class of objects, and are in the interval:
$0.55 \lapp \alpha_{ro} \lapp 0.89$.
 
Finally, recent HST observations of the M87 jet by
Sparks et al. (1996) allowed the authors to estimate the value of
the spectral index in the optical-UV band, in the
range of frequencies observed by the FOC detector on board HST
($5.6 \times 10^{14} \le \nu \le 2 \times 10^{15}$ Hz),
$\alpha_{o-UV} \sim 1.3$.
In these frequency intervals our values span from $\alpha_{opt,o-UV} \sim 0.51$
to $\sim 1.13$.

\begin{table}[hbt]

\caption{Average spectral indexes for the 10 parcels in radio,
radio to optical and optical-UV frequency ranges}

\begin{center}
\begin{tabular}{|c|c|c|c|}  \hline

particle&radio&radio-opt&opt-UV     \\
\hline
1    & 0.49 & 0.71 & 0.96  \\
2    & 0.67 & 0.89 & 1.13  \\
3    & 0.65 & 0.74 & 0.83  \\
4    & 0.48 & 0.49 & 0.51  \\
5    & 0.65 & 0.65 & 0.65  \\
6    & 0.43 & 0.55 & 0.76  \\
7    & 0.50 & 0.58 & 0.71  \\
8    & 0.50 & 0.70 & 0.83  \\
9    & 0.59 & 0.82 & 1.00  \\
10   & 0.54 & 0.61 & 0.71  \\
\hline
\end{tabular}
\end{center}
 \end{table}

In Fig. 5 we show the values of the radio spectral index
calculated at each shock for each of the 10 Lagrangian
particles. The dotted lines
define the time intervals used in Table 4.
From the figure it is clear how the spectral index starts from an
initially high value (here the first linear phase is not represented,
and so the values derived from the values of $\gamma$ at $t=12$ in Table 3
are not shown), and it decreases
as the particles enter a higher and higher number of shocks, during
the acoustic phase of the instability.

When the jet enters the mixing phase, shocks become rare and weak,
and the spectral index increases. 
This is an interesting point, because the final increase in the
spectral index is a feature which comes out from the evolution of the
instability, and does not depend on the initial parameters or assumptions
of our model; on the contrary, assuming a different initial slope for the
electrons spectrum could give a different trend for the spectral index in
the shock-dominated region of the jet: for example, if the electrons were
injected directly at the beginning of the acoustic phase, they would not
undergo the heavy losses and steepening as it happens in the first linear
phase of the evolution; a different trend could also be obtained abandoning
the assumption of Lagrangian advection of the particles by the fluid,
or if, following a different choice of the initial jet parameters, many
more shocks were involved or shocks with stronger
compression ratios. In this way, the decreasing trend would be probably
limited to the first section of the knots chain, and would be followed
by a region of quasi-constant or increasing spectral index.

This behavior is in agreement with the trend in the radio spectral 
index observed in a number of extragalactic radio jets: for example, 
Sparks et al. (1996) and  Meisenheimer et al. (1996) show that 
in M87 the spectral index oscillates 
around the mean value for the first section of the knot chain, 
and increases for the  farthest knots from the source, and this is 
a feature common to all the frequency ranges that were studied;
in NGC 6251  the spectral index is   almost constant along the jet,
with oscillations around the mean value $\alpha \sim 0.6$ , 
and it shows a gradual 
steepening to 0.8 shortly before the jet widens up and mixes with 
the diffuse lobe emission (Mack et al. 1998)
and Perley et al. (1984); similar values and trends are also observed
in Cen A (Burns et al. 1983), 3C 279 (De Pater \& Perley 1983)
and 3C 31 (Strom et al. 1983).

Fig. 5 and Table 3 show that the value reached by the spectral
index is not strictly correlated to the initial radial position of the
Lagrangian particle in the jet: a small number of close strong shocks can
flatten the electrons distribution function in the same way as a
higher number of distributed shocks. The particles initially located
nearer to the jet axis, nevertheless, show a quasi-constant spectral
index for a longer time: compare for example particles 1,2,6,7,8
which cross a high number of shocks, to particles 4,5,9,10 whose spectrum
is flat instead only for a limited interval of time.

\begin{figure}[htbp]

{\includegraphics[width=\hsize,height=14.cm,bb=100 280 470 710]{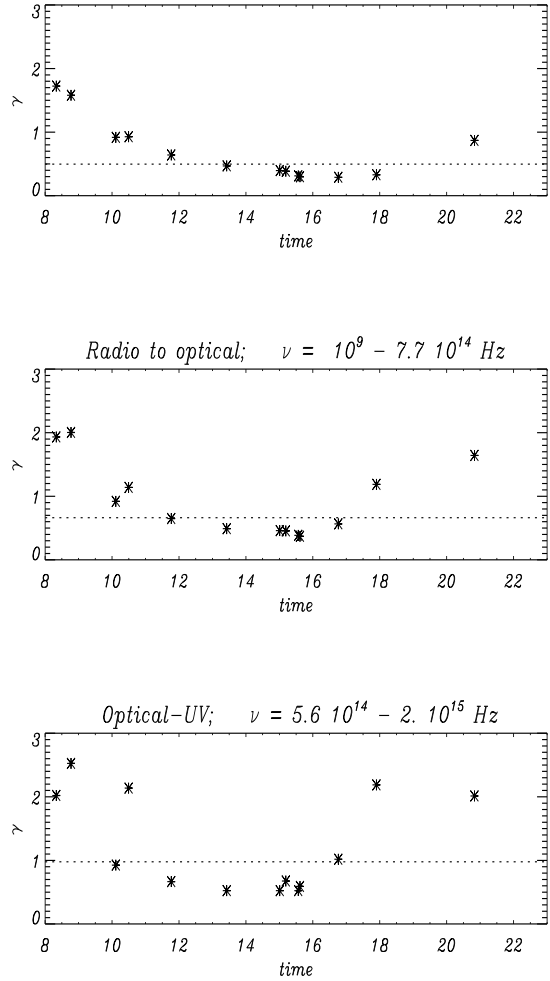}}

\caption{Radio, radio to optical and optical to UV post-shock spectral 
indexes calculated for particle 1. The dotted lines show, as in Figure 5,
the average values calculated in the time interval $9.5 \lapp t \lapp
18.5$, corresponding to the shock-dominated stage of the evolution
 }
   \end{figure}

In Fig. 6 the values and evolution of the spectral index $\alpha$
associated to
particle 1 in the different frequency ranges are compared. Energetic
electrons undergo synchrotron losses more rapidly, and the high energy
section of the spectrum steepen more quickly. This leads to a higher
average value of the spectral index in the optical-UV region respect to
the radio value and a shorter temporal (and thus spatial) extent of
the region where the spectrum is flatter;
$\alpha_{o-UV}$ starts to grow already before the onset of the mixing
phase.

Fig. 7a shows the behavior of the spectral index $\alpha$ for the
post-shock distribution function (stars) as well as for the inter-knot
regions (continuous line). In Fig. 7a the temporal evolution has been
translated into a spatial evolution, and the values of the spectral index
are plotted as a function of the distance of the particle from the
injection point. The trend in Fig. 7a reproduces qualitatively
the behavior of the
optical and radio-optical spectral index observed by
Meisenheimer at al. (1996) for M87, although in our model the
oscillations between the values in knot and inter-knot regions have a
bigger amplitude at the beginning of the shock chain, while just before
the onset of the mixing phase the pattern is almost constant;
 in M87, which is the only object which can be observed with 
enough resolution to provide these data,
 the knot to inter-knot differences
are always smaller than $\Delta\alpha \sim 0.2$, while in our case
the average value is $\Delta\alpha \sim 0.23$. 

Fig. 7 only shows the section of the jet where
shocks are present. The first section, where the
perturbation grows according to the linear theory, corresponds to the
{\it gap} between the source and the first bright knot observed in
most astrophysical jets. If we adopt for the jet radius an average
value estimated for a typical radio jet (see Table 2)
we find consistent  values for the inter-shock distances between our
model and the inter-knot distances that are generally observed, while
the length of the emission gap in our model is an order of magnitude
larger than observed. However, we recall that the simple cylindrical
geometry adopted here forces us to select  an average, constant value for
the jet radius (and for all the physical parameters in general),
while it is well known that in the inner regions the jet radius 
can be up to two orders of magnitude smaller
(in the case of M87, for example, the jet radius is below the
HST 0.03$^{\prime \prime}$ resolution (i.e. $\lapp 2.5$ pc) close to
the source, Biretta (1998),
and increases with distance with an opening angle $5-7^{\circ}$
at VLA wavelengths). Bearing this in mind,
a gap of $\sim$ 50 jet radii from the source, where the first shock
forms, would correspond to a distance consistent with
the gaps that are currently observed in radio jets.

\begin{figure}[htbp]

{\includegraphics[width=\hsize,height=9.cm,bb=100 60 570 640]{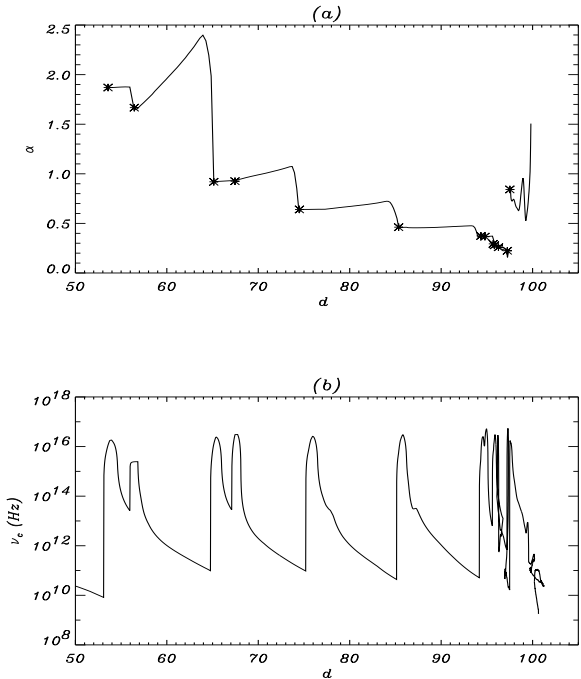}}

\caption{Panel (a): variation in the spectral index values with distance from the
jet injection point, for the electron distribution function associated
with particle 1. The stars represent the post-shock values of the spectral
index. Panel (b): Cutoff frequency in the electron distribution
function as a function of the distance from the jet injection point.
 }
   \end{figure}

Finally, Fig. 7b shows the spatial trend of the cutoff frequency
$\nu_c = 4.3 \times 10^6 \ B_\perp \Gamma^2 $ Hz
as it results from the evolution of the cutoff energy along the jet
for particle 1. The values of the peaks in Fig. 7b are
model-dependent, 
i.e. they depend on the maximum value that the electron
energy can attain in our model since
 we assume that at shocks the electrons are
re-accelerated up to the initial maximum energy value
$\Gamma_{\rm max} = 100 \ \Gamma_0 \sim 2.55 \times 10^7$.
This assumption only influences the peak values: other simulations
performed assuming different values for the maximum allowed energy
gave an exactly similar trend, with different peak heights (but same
widths).

Our results imply that while the knots regions are bright at least up
to optical frequencies, in the inter-knot regions only radio emission
can take place. 
Therefore, as long as the cutoff frequency is concerned,
our simplified model does not explain the features observed in the 
M87 jet, where Meisenheimer et al. (1996) find that the cutoff 
frequency in the inter-knot regions lies in the optical region
of the spectrum. 
The authors report that the cutoff frequency changes by no more than a
factor 3 along the jet. Even if we smooth our data at the resolution
of Meisenheimer et al. we still find variations in $\nu_c$ of two
orders of magnitude, and $\nu_c \le 10^{12}$ in the inter-knot regions,
rendering them optically invisible.

\section{Conclusions}

We presented a self-consistent study of particles
shock acceleration, in the test particle approximation,
in an astrophysical jet.

We found that the non linear growth of the Kelvin-Helmholtz instability
is able to produce a high number ($\sim 10$) of strong shocks
where the energy of a distribution of relativistic electrons advected by the
fluid can be increased via the first order Fermi-like mechanism of
diffusive shock acceleration. The acceleration process is efficient enough
to compete with the losses via synchrotron radiation and adiabatic
expansion taking place in the whole body of the jet; the flattening
in the energy distribution function of the electrons gives rise to
a flat frequency spectrum of the synchrotron radiation emitted
which can be directly compared with observations.

We discussed the acceleration process connected to the subsequent phases
of the instability evolution; involving 10 Lagrangian parcels
of the same initial distribution function located at different
initial positions allowed us to find that the acceleration process is not
dependent on the initial distance of the particles from the jet axis.
Moreover we found that the total number of relativistic electrons
associated to each Lagrangian particle is conserved during the whole
evolution, up to the final stages when shocks cannot form any more and
the electrons simply lose energy becoming non-relativistic.

A further improvement of our model would be the adoption of a numerical
code written to solve the full magnetohydrodynamical equations: this
would give the proper evolution of the magnetic field, whose components
in our calculations are simply advected by the fluid. A situation
in which the force exerted by the magnetic field on the fluid is taken
into account could probably lead to a reduction of the shock strength,
but it might also imply a longer duration of the shock-dominated
phase of the instability evolution. The problems arising in the
development of a reliable multidimensional MHD code have up to now
limited the numerical study of magnetized jets.

From our results we were able to evaluate the average values for the
spectral index of the emitted synchrotron radiation in different
frequency ranges, obtaining values consistent with those derived from
observations of extragalactic jets.
We also gave estimates of the cutoff frequency expected in the knots
and in the inter-knot regions of the jet.

Although the main aim of this work was to give a physical description
of how the diffusive acceleration mechanism can work in a
Kelvin-Helmholtz unstable jet, we could also derive
the trends of the above described observational quantities
with the position along the jet axis, and this was
achieved translating our results in time
into behavior with space. It is necessary to keep it into account
when we refer for example to Fig. 7: the plots in Fig. 7a and 7b
are not snapshots of the jet at a certain time, but are obtained
following a parcel of the jet material as it moves downstream, thus
evolving both in space and in time.
For a more realistic application to observational data of the 
mechanism described in this paper,
a `spatial' analysis of the evolution of a jet propagating
in a medium with decreasing density is required, in order to obtain
snapshots of the whole (non-cylindrical)
jet at selected times; this implies the adoption
of large grids (see for example Micono et al., 1998) and consequently
of a large number of Lagrangian particles, to cover with a good
statistic the whole jet body; 
 in this way we could also obtain indications on the variation
of the density of the Lagrangian particles in different jet zones,
and thus on their relative brightness.
This study will be the subject of a
future work.

\section{References}

\noindent
Achterberg, A., 1990, {\it Physical Processes in Hot Cosmic
Plasmas}, Kluwer Academic Publishers, 67

\noindent
Bell, A.L., 1978, {\it M.N.R.A.S.} {\bf 182},
147

\noindent
Bicknell, G.V., Begelman, M.C., 1996, {\it ApJ} {\bf 467}, 597

\noindent
Biretta, J.A., Zhou, F., Owen, F.N., 1995, {\it ApJ} {\bf 447}, 582

\noindent
Birkinshaw, M., 1991, {\it in Beams and Jets in Astrophysics},
P.A. Hughes ed., Cambridge University Press, 278

\noindent
Bisnovatyi-Kogan, G.S., Lovelace, R.V.E., 1995,  {\it A\&A} {\bf 296}, L19

\noindent
Blandford, R.D., and Eicheler, D., 1987, {\it Physics
Reports} {\bf 154}, 1

\noindent
Blandford, R.D., and Ostriker, J.P., 1980, {\it ApJ} {\bf 237}, 793

\noindent
Bodo, G., Massaglia, S., Ferrari, A.,  Trussoni, E., 1994,
{\it A\&A} {\bf 283}, 655

\noindent
Bridle, A.H., Perley, R.A., 1984, {\it A.R.A.A.} {\bf 22}, 319

\noindent
Burns, J.O., Feigelson, E.D., Schreier, E.J., 1983, {\it Ap.J.} 
{\bf 273}, 128

\noindent
Colella, P., Woodward, P.R., 1984, {\it J.Comp.Phys.} {\bf 54}, 174

\noindent
De Pater, I., Perley, R.A., 1983, {\it Ap.J.} {\bf 273}, 64

\noindent
Drury, L. O'C. 1983, {\it Rep. prog. Phys.} {\bf 46}, 973

\noindent
Ferrari, A., Trussoni, E., Zaninetti, L., 1978, {\it A\&A} {\bf 64},
43

\noindent
Ferrari, A., Melrose, D.B. 1997, {\it Mem. Soc. Astron. Ital.} {\bf 68}, 171

\noindent
Fraix-Burnet, D., Golombeck, D., Macchetto, F.D., Nieto, J.L., 
Lelievre, G., Perryman, M.A.C., Di Serego Alighieri, S., 1991a,
{\it AJ} {\bf 101}, 88,

\noindent
Fraix-Burnet, D., Golombeck, D., Macchetto, F.D., 1991, {\it AJ} 
{\bf 102}, 562

\noindent
Hardee P.E., Norman M.L., 1988a, {\it ApJ} {\bf 334}, 70

\noindent
Hardee P.E., Norman M.L., 1988b, {\it ApJ} {\bf 334}, 80

\noindent
Heavens, A.F., Drury, L.O'C., 1988, {\it M.N.R.A.S.} {\bf 235}, 997

\noindent
Jones, T.W., Ryu, D., Engel, A., 1999, {\it ApJ} {\bf 512}, 105

\noindent
Kardashev N.S., 1962, {\it Soviet Astron.} {\bf 6}, 317

\noindent
Kirk, J.G., Schneider, P., 1987, {\it ApJ} {\bf 315}, 425

\noindent
Kirk, J.G., 1994,
{\it Plasma Astrophysics}, Saas-Fee Advanced Course 24
(Lecture Notes 1994), Springer--Verlag, Berlin, 225

\noindent
Lesch, H., Birk, G.T., 1998, {\it ApJ} {\bf 499}, 167

\noindent
Li, H., Miller, J.A., Colgate, S.A., 1997, {\it Relativistic
Jets in AGNs}, Proceedings of the International Conference, 162

\noindent
Mack, K.H., Klein, U., O'Dea, C.P., Willis, A.G., Saripalli, L., 1998,
{\it A\& A} {\bf 329}, 442

\noindent
Massaglia, S., Bodo, G., Ferrari, A., Rossi, P., 1996, {\it Jets from Stars
and Galactic Nuclei}, Proceedings of a Workshop Held at Bad Honnef
(3-7 July 1995), 275 

\noindent 
Matthews, A.P., Scheuer, P.A.G., 1990, {\it M.N.R.A.S.} {\bf 242}, 616

\noindent 
Matthews, A.P., Scheuer, P.A.G., 1990, {\it M.N.R.A.S.} {\bf 242}, 623

\noindent
Meisenheimer, R\"oser, H.-J., Schl\"otelburg, M.,
1996, {\it Jets from Stars
and Galactic Nuclei}, Proceedings of a Workshop Held at Bad Honnef
(3-7 July 1995), 231

\noindent
Meisenheimer, K., Neumann, M., R\"oser, H.-J.,
1996, {\it A\&A} {\bf 307}, 61

\noindent
Meisenheimer, K., Neumann, M., R\"oser, H.-J.,
1996, {\it A\&A} {\bf 307}, 61

\noindent
Melrose, D.B., Pope, M.H., 1993, {\it Proc. Astron. Soc. Australia}
{\bf 10}, 222

\noindent
Micono, M., Massaglia, S.,  Bodo G., Rossi, P., Ferrari, A., 1998,
{\it A\&A} {\bf 333}, 989

\noindent
Ostrowsky, M., 1990,  {\it A\&A} {\bf 238}, 435

\noindent
Ostrowsky, M., 1998,  {\it A\&A} {\bf 335}, 134

\noindent
Perley, R.A., Bridle, A.H., Willis, A.G., 1984, {\it Ap.J.Suppl} 
{\bf 54}, 291

\noindent
Reid, M.J., Biretta, J.A., Junor, W., Muxlow, T.W.B., Spencer, R.E.,
1989, {\it ApJ} {\bf 336}, 112

\noindent
Rossi, P., Bodo, G., Massaglia, S., Ferrari, A., 1997, {\it A\&A}
{\bf 321}, 672

\noindent
Schneider, P., 1993, {\it A\&A} {\bf 278}, 315

\noindent
Sparks, W.B., Biretta, J.A., Macchetto, F., 1996, {\it ApJ}
{\bf 473}, 254

\noindent
Stone, J.M., Xu, J., Hardee, P.E., 1997, {\it ApJ} {\bf 483}, 136

\noindent
Strom, R.G., Fanti, R., Parma, P., Ekers, R.D., 1983, 
{\it A. \& A.} {\bf 122}, 305

\noindent
White, R.L., 1985, {\it ApJ} {\bf 289}, 698

\noindent
Zensus, J.A., Unwin, S.C., Cohen, M.H., Biretta, J.A., 1990, {\it A.J.}
{\bf 100}, 1777

\end{document}